\documentclass[12pt]{article}
\usepackage{graphicx}
\usepackage{epsfig}
\def\lsim{\raise0.3ex\hbox{$<$\kern-0.75em\raise-1.1ex\hbox{$\sim$}}}
\def\gsim{\raise0.3ex\hbox{$>$\kern-0.75em\raise-1.1ex\hbox{$\sim$}}}

\def\mean#1{\left<#1\right>}
\def\Journal#1#2#3#4{{#1}{\bf #2} (#4) #3}

\def\IJMPE{{Int. J. Mod. Phys. E}}

\def\NPA{{Nucl. Phys. A}}
\def\NPB{{Nucl. Phys. B}}
\def\PLB{{Phys. Lett. B}}
\def\PLC{Phys. Repts.\ }

\def\PRL{Phys. Rev. Lett.\ }
\def\PRD{{Phys. Rev. D}}
\def\PRC{{Phys. Rev. C}}

\def\ARNPS{{Ann. Rev. Nucl. Part. Sci.\ }}

\def\RPP{Rep. Prog. Phys.\ }
\normalsize
\begin{document}
\title{Results from PHENIX at RHIC}
\author{M.~J.~Tannenbaum 
\thanks{Research supported by U.S. Department of Energy, DE-AC02-98CH10886.}
\\ Physics Department, 510c,\\
Brookhaven National Laboratory,\\
Upton, NY 11973-5000, USA\\
mjt@bnl.gov}
\maketitle
\section{Introduction} 
In nucleus-nucleus collisions at large c.m. energy or baryon density, a phase transition is expected from a state of nucleons containing confined quarks and gluons to a state of ``deconfined'' (from their individual nucleons) quarks and gluons, in chemical and thermal equilibrium, covering a volume that is many units of the confining length scale. This state of nuclear matter was originally given the name Quark Gluon Plasma(QGP)~\cite{Shuryak80}, a plasma being an ionized gas. However the results at RHIC~\cite{MJTROP} indicated that instead of behaving like a gas of free quarks and gluons, the matter created in heavy ion collisions at nucleon-nucleon c.m. energy $\sqrt{s_{NN}}=200$ GeV appears to be more like a {\em liquid}. This matter interacts much more strongly than originally expected, as elaborated in recent peer reviewed articles by the 4 RHIC experiments~\cite{EXWP}, which inspired the theorists~\cite{THWPs} to give it the new name ``sQGP" (strongly interacting QGP).  

\section{Issues in Relativistic Heavy Ion Physics}

   Since 1986, the `gold-plated' signature of deconfinement was thought to be $J/\Psi$ suppression. Matsui and Satz~\cite{MatsuiSatz86} proposed that $J/\Psi$ production in A+A collisions will be suppressed by Debye screening of the quark
color charge in the QGP which would dissolve the $c, \bar c$ bound state. ``Anomalous suppression'' of $J/\Psi$ was found in Pb+Pb collisions at the CERN SpS $\sqrt{s_{NN}}=17.2$ GeV~\cite{NA50PLB450}. The search for $J/\Psi$ suppression and thermal photon/dilepton radiation from the QGP drove the design of the RHIC experiments. 
\begin{figure}[!hbt]
\begin{center}
\begin{tabular}{cc}
\psfig{file=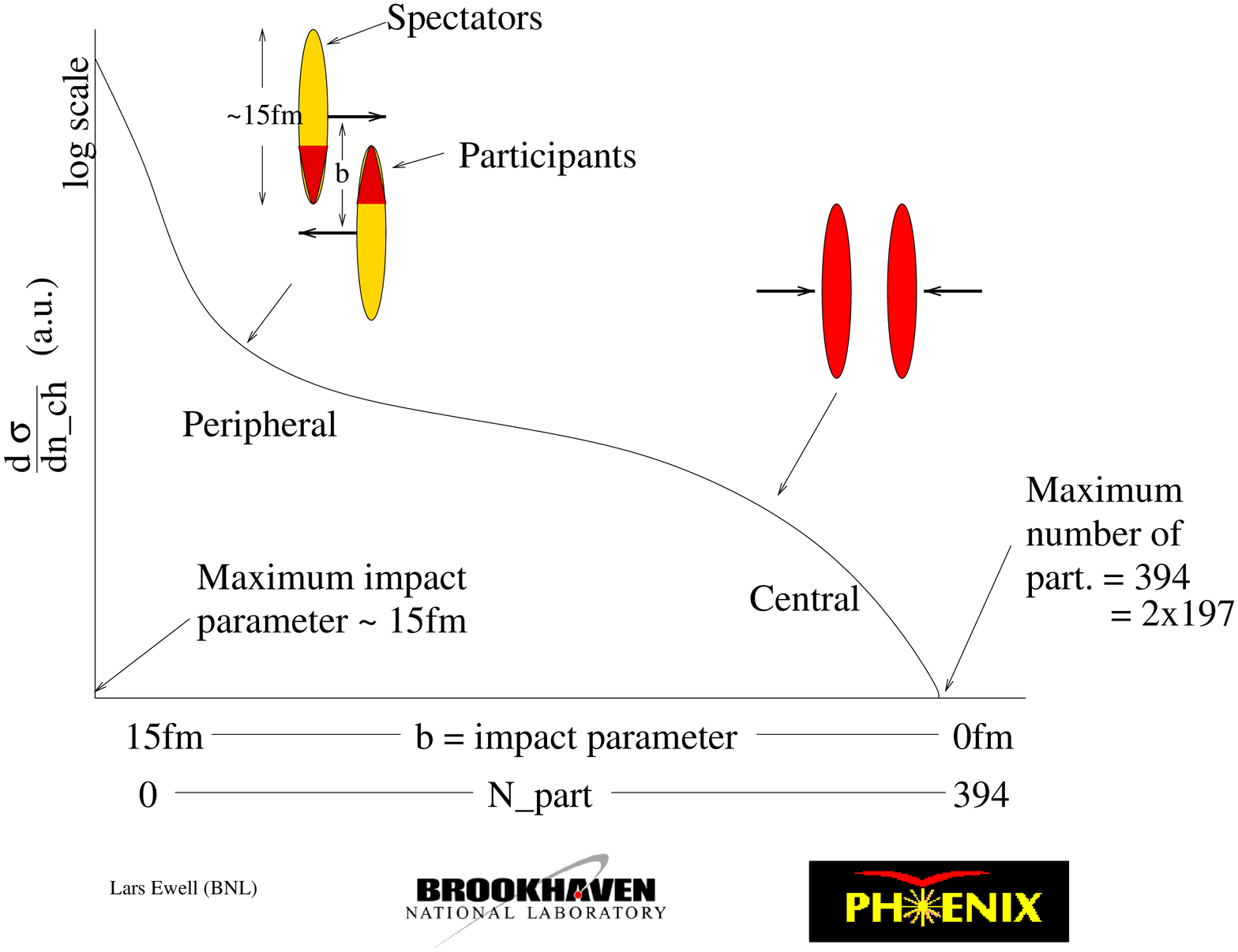,height=0.50\linewidth,angle=-90,bbllx=10, bblly=14, bburx=533, bbury=778,clip=}&
\psfig{file=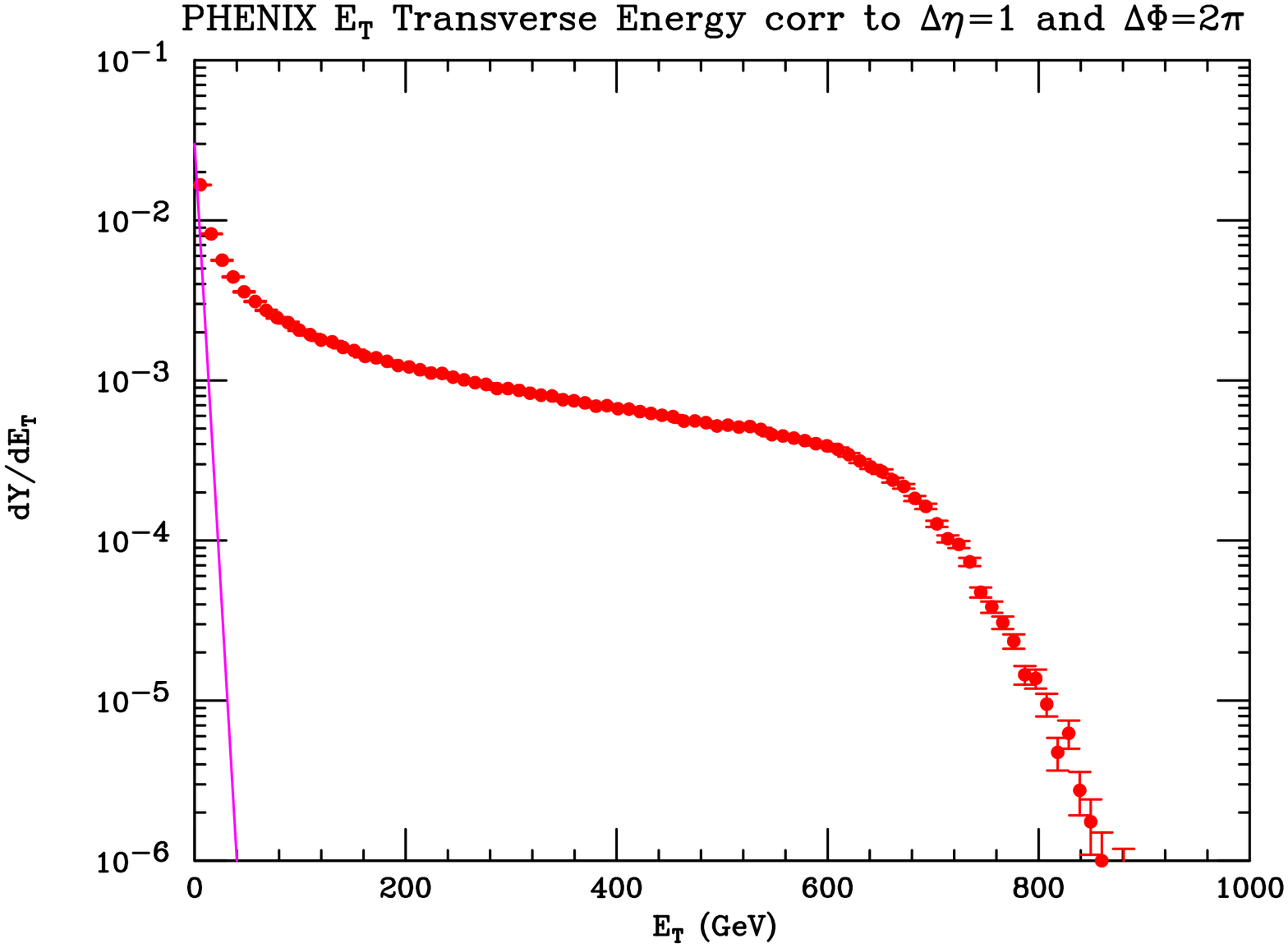,height=0.45\linewidth,angle=-90}\end{tabular}
\end{center}\vspace*{-0.15in}
\caption[]{a) (left) Schematic of collision of two nuclei with impact parameter $b$. The curve with the ordinate labeled $d\sigma/d n_{\rm ch}$ represents the relative probability of charged particle  multiplicity $n_{\rm ch}$ which is directly proportional to the number of participating nucleons, $N_{\rm part}$. b)(right) Transverse energy ($E_T$) distribution in Au+Au and p-p collisions at $\sqrt{s_{NN}}=200$ GeV from PHENIX~\cite{ppg019}.  
\label{fig:nuclcoll}}
\end{figure}

\begin{figure}[!bht]
\begin{center}
\begin{tabular}{cc}
\psfig{file=figs/STARJet+AuAu-g.epsf,width=0.64\linewidth}&\hspace*{-0.025\linewidth}
\psfig{file=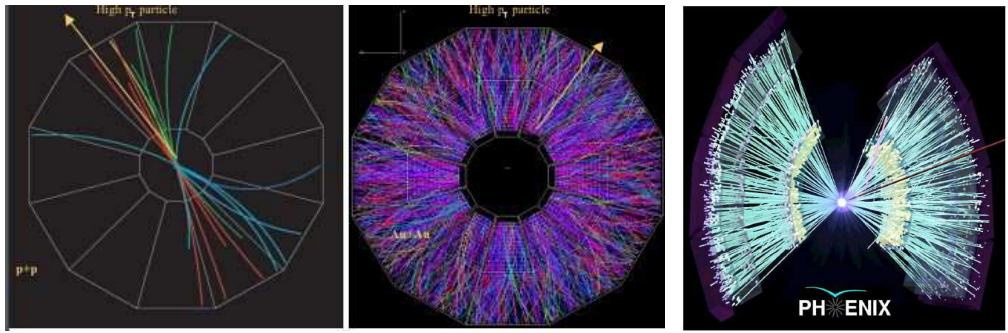,width=0.315\linewidth,height=0.315\linewidth}
\end{tabular}
\end{center}\vspace*{-0.15in}
\caption[]{a) (left) A p-p collision in the STAR detector viewed along the collision axis; b) (center) Au+Au central collision at $\sqrt{s_{NN}}=200$ GeV in the STAR detector;  c) (right) Au+Au central collision at $\sqrt{s_{NN}}=200$ GeV in the PHENIX detector.  
\label{fig:collstar}}

\end{figure}

	Another main concern of experimental design in RHI collisions is the huge multiplicity in A+A central collisions compared to  p-p collisions. 
A schematic drawing of a collision of two relativistic Au nuclei is shown in Fig.~\ref{fig:nuclcoll}. In the center of mass system of the nucleus-nucleus collision, the two Lorentz-contracted nuclei of radius $R$ approach each other with impact parameter $b$. In the region of overlap, the ``participating" nucleons interact with each other, while in the non-overlap region, the ``spectator" nucleons simply continue on their original trajectories. The degree of overlap is called the centrality of the collision. The energy of the inelastic collision is predominantly dissipated by multiple particle production, where $n_{\rm ch}$, the number of charged particles produced, is directly proportional to the number of participating nucleons ($N_{\rm part}$) as sketched on Fig.~\ref{fig:nuclcoll}. Thus, $n_{\rm ch}$ or the total transverse energy $E_T$ in central Au+Au collisions is roughly $A$ times larger than in a p-p collision, as shown in the measured transverse energy spectrum in the PHENIX detector for Au+Au compared to p-p (Fig.~\ref{fig:nuclcoll}b)~\cite{ppg019} and in actual events from the STAR and PHENIX detectors at RHIC in Fig.~\ref{fig:collstar}.

	As it is a daunting task to reconstruct all the particles produced in such events, the initial experiments at RHIC concentrated on the measurement of single-particle or multi-particle inclusive variables to analyze RHI collisions, with inspiration from the CERN ISR which emphasized those techniques before the era of jet reconstruction. The two major detectors presently in operation at RHIC (Fig.~\ref{fig:collstar}), are STAR, a conventional TPC, and PHENIX, a very high granularity high resolution special purpose detector covering a smaller solid angle designed to measure and trigger on rare processes involving leptons, photons and identified hadrons at the highest luminosity.

	In addition to the large multiplicity, there are two other issues in A+A  physics which are different from p-p physics: i) space-time issues, both in momentum space and coordinate space---for instance what is the spatial extent of fragmentation? is there a formation time/distance?; ii) huge azimuthal anisotropies of particle production in non-central collisions (colloquially collective flow) which are interesting in their own right but can be troublesome. 
	
\begin{figure}[!thb]
   \begin{center}
\includegraphics[width=0.45\linewidth]{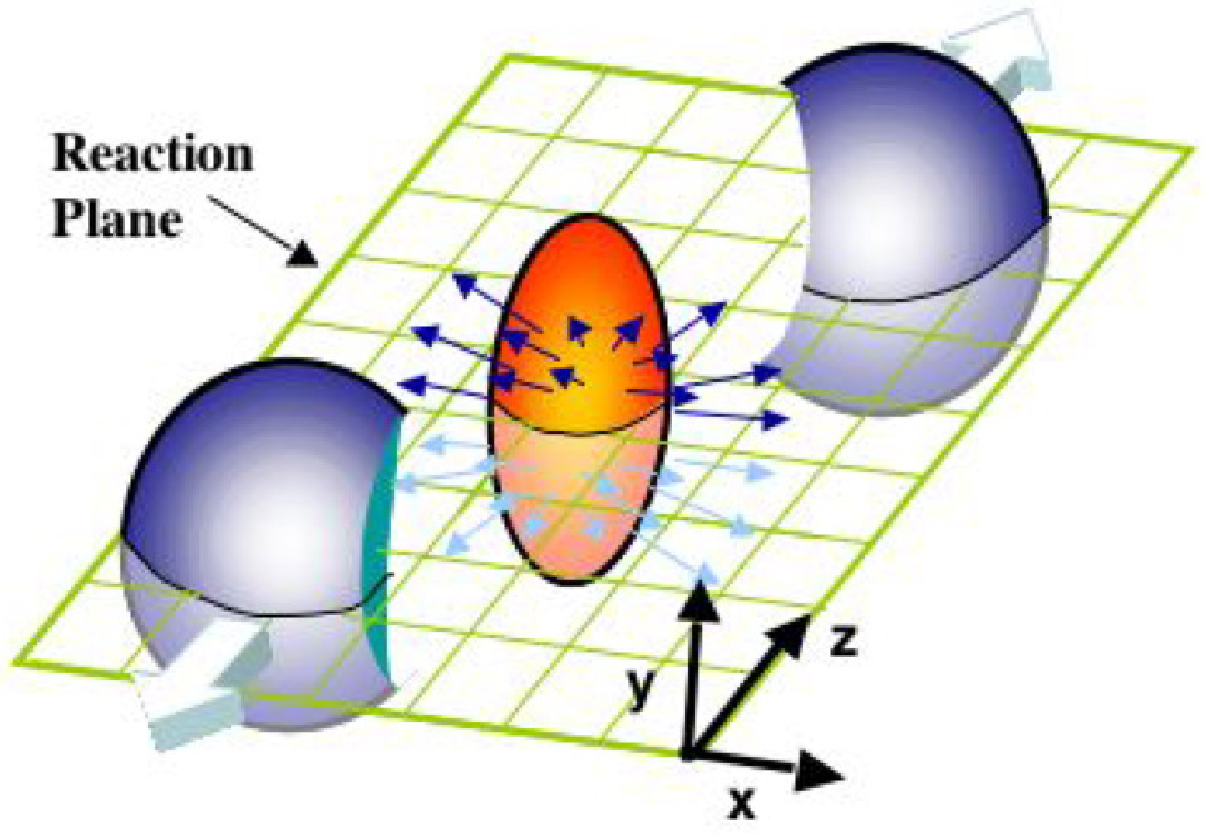}
\includegraphics[width=0.54\linewidth]{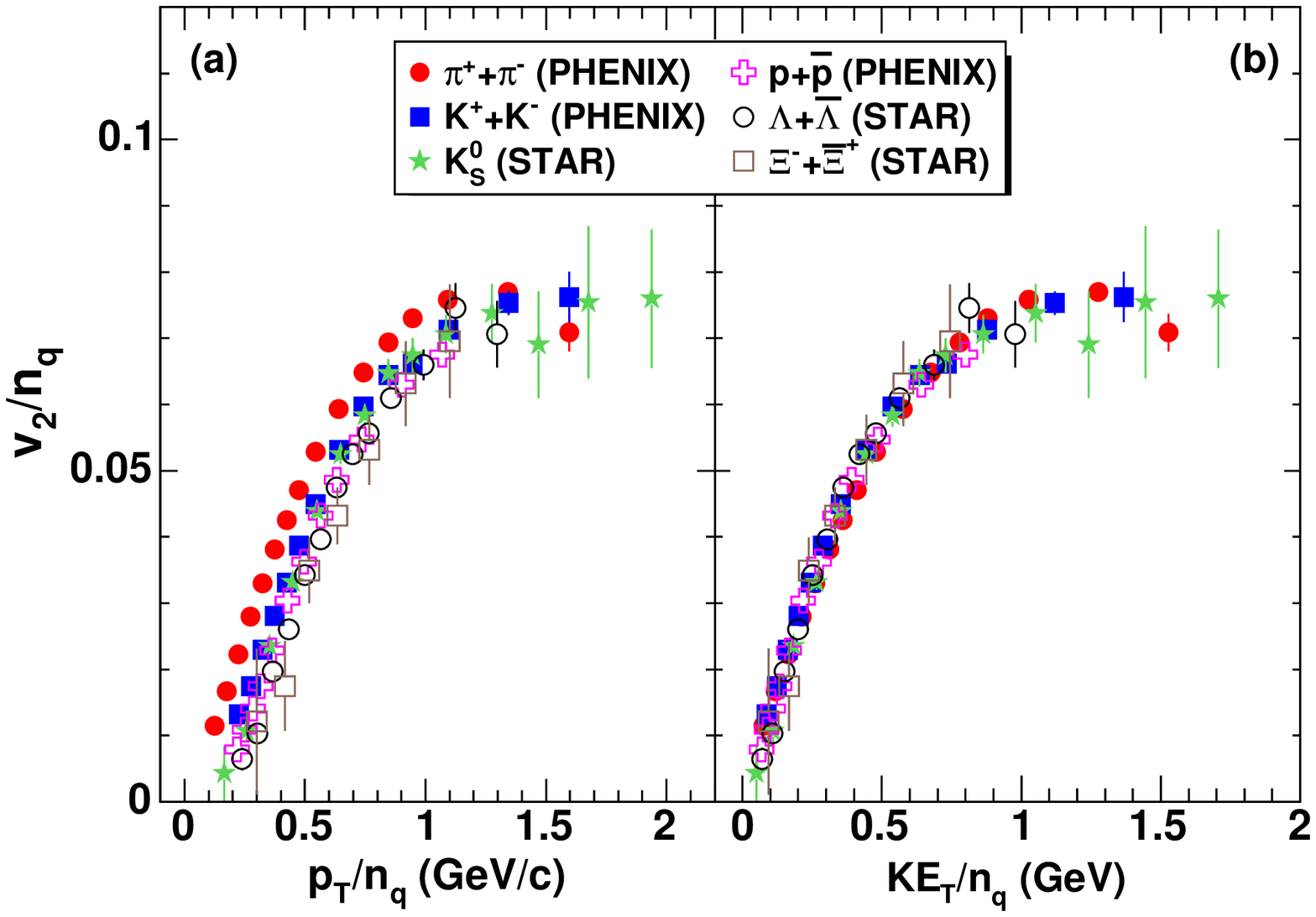}
\end{center}\vspace*{-0.25in}
\caption[]{(left) Almond shaped overlap zone generated just after an A+A collision where the incident nuclei are moving along the $\pm z$ axis. The reaction plane is defined by the $z$ axis and the $x$ axis (which contains the impact parameter vector and defines $\Phi_R=0$). (right) Measurements of elliptical-flow ($v_2$)  for identified hadrons plotted as $v_2$ divided by the number of constituent quarks $n_q$ in the hadron as a function of (a) $p_T/n_q$, (b) $KE_T/n_q$~\cite{PXArkadyQM06}.   
\label{fig:MasashiFlow1}}
\end{figure}

	Collective flow, or simply flow, is a collective effect which can not be obtained from a superposition of independent N-N collisions. The almond shaped overlap region of the A+A collision causes the particles to be emitted more favorably in the reaction plane (see Fig.~\ref{fig:MasashiFlow1}).   
The semi-inclusive single particle spectrum is modified by an expansion in harmonics of the azimuthal angle of the particle with respect to the reaction plane, $\phi-\Phi_R$:   
\begin{equation}
{Ed^3 N \over dp^3}={d^3 N\over p_T dp_T dy d\phi}
={d^3 N\over 2\pi\, p_T dp_T dy} \left[ 1+\sum_n 2 v_n \cos n(\phi-\Phi_R)\right] .
\label{eq:siginv2} 
\end{equation} 
The expansion parameter $v_2$, called elliptical flow, is predominant at mid-rapidity and shows scaling rules which imply that it is the constituent quarks rather than the final hadrons that flow (Fig.~\ref{fig:MasashiFlow1}b)~\cite{PXArkadyQM06}.

\section{Measurements in proton-proton collisions}
          \begin{figure}[!th]
\begin{center}
\includegraphics[width=0.48\linewidth]{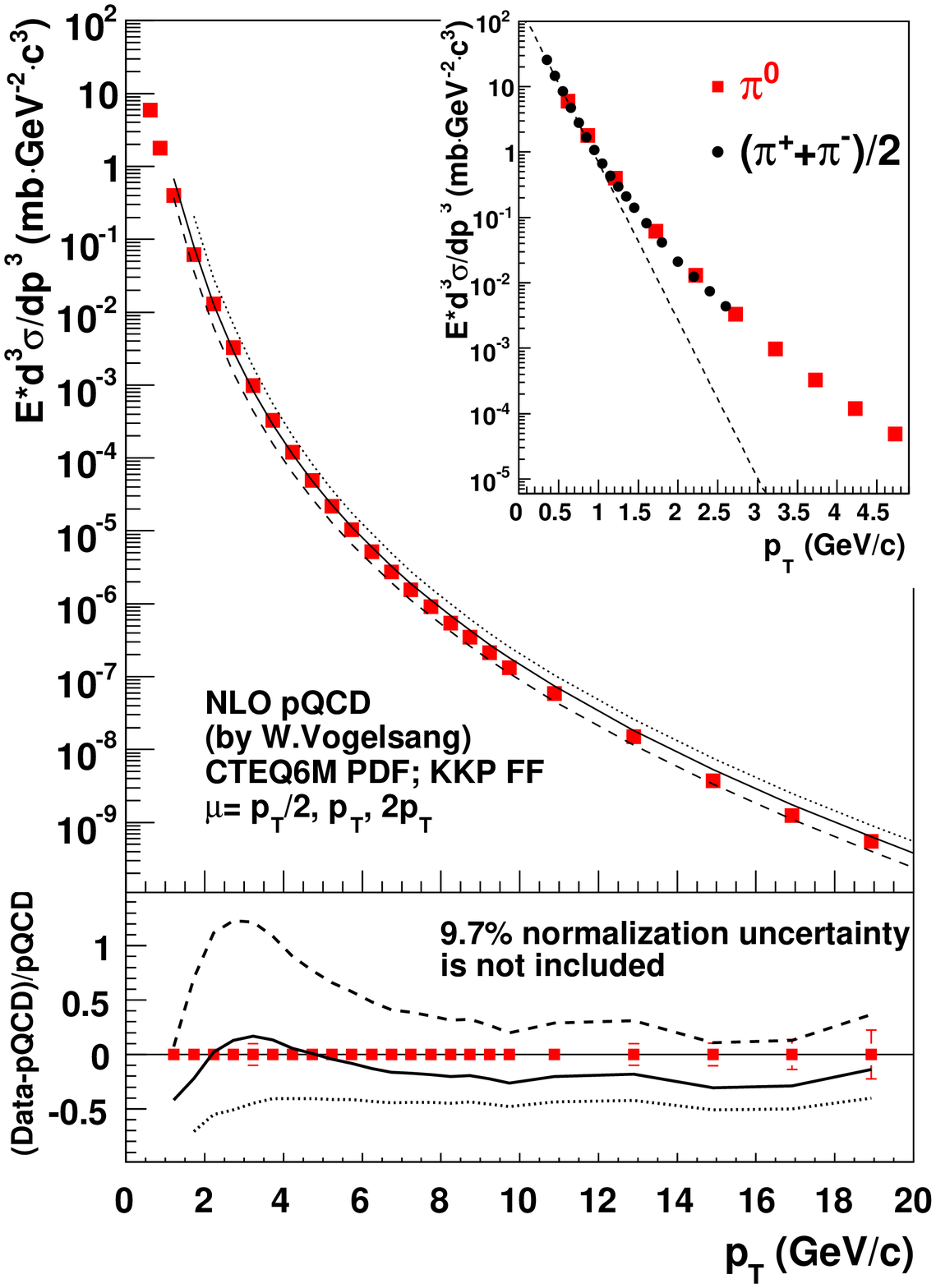}
\hspace*{-0.03\linewidth}\includegraphics[width=0.53\linewidth]{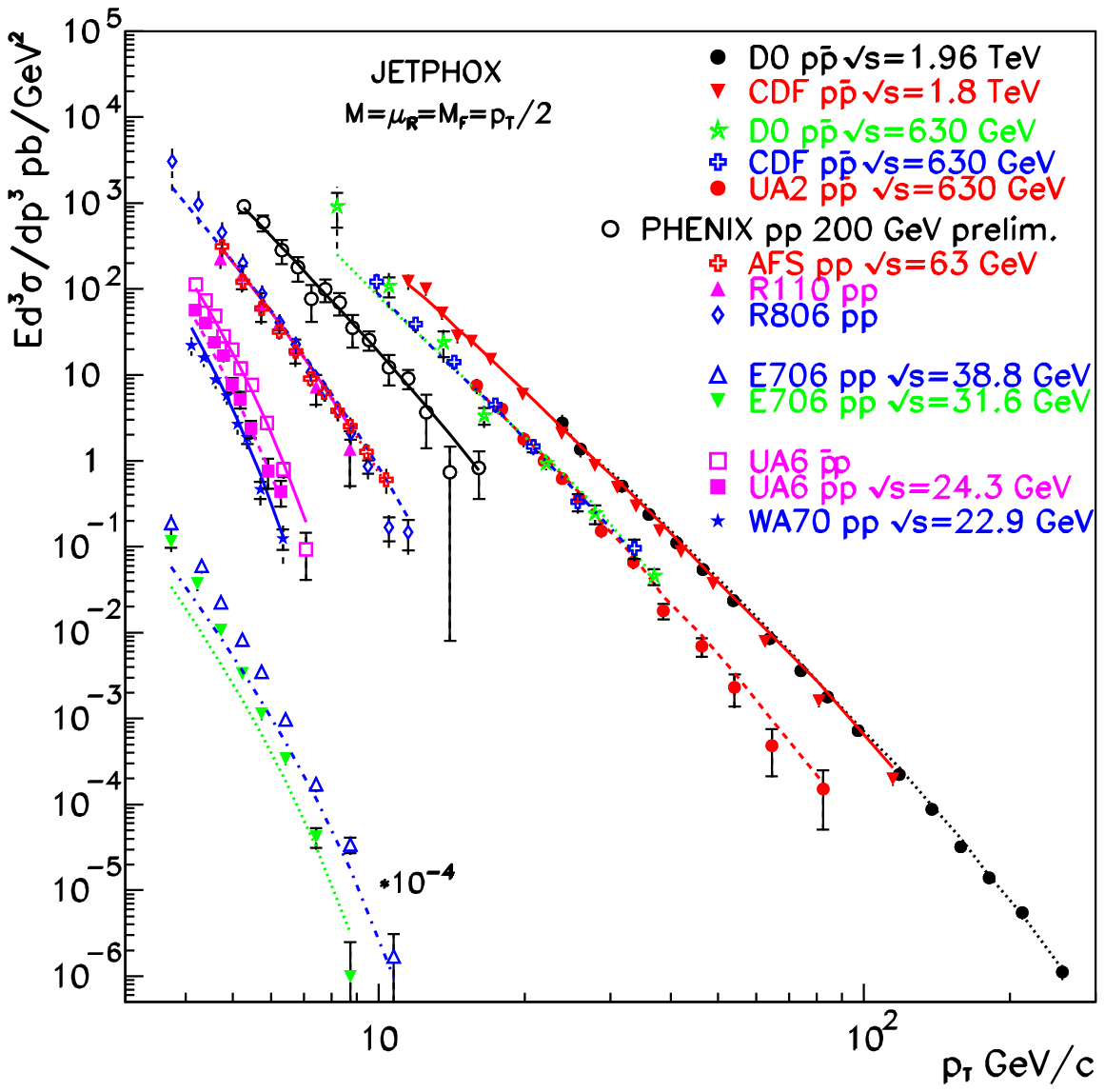}
\end{center}
\caption[]{a) (left) PHENIX measurement of invariant cross section of $\pi^0$ vs. $p_T$ at mid-rapidity in p-p collisons at $\sqrt{s}=200$ GeV.~\cite{PXpi0PRD}. b) (right) PHENIX measurement of inclusive direct-single $\gamma$ in p-p collisons at $\sqrt{s}=200$ GeV ($\circ$), together with all previous data compared to the theory.~\cite{Aurenche} }
\label{fig:pi0GamRHIC}
\end{figure} 
     In order to understand whether an effect observed in A+A collisions exhibits a sensitivity to collective effects or to the presence of a medium such as the QGP it is important to establish a precise baseline measurement in p-p collisions at the same value of nucleon-nucleon c.m. energy $\sqrt{s_{NN}}$. PHENIX measurements of the invariant cross section, $E d^3\sigma/dp^3$, for $\pi^0$ and direct-single-$\gamma$ production in p-p collisions at $\sqrt{s}=200$ GeV are shown in Fig.~\ref{fig:pi0GamRHIC}a and Fig.~\ref{fig:pi0GamRHIC}b, respectively. The inset on Fig.~\ref{fig:pi0GamRHIC}a shows that the $\pi^0$ cross section is exponential $\sim e^{-6p_T}$ for $p_T< 2$ GeV/c, which is the region of soft-multiparticle physics. For $p_T > 2$ GeV/c the spectrum is a power law which is indicative of the hard-scattering of the quark and gluon constituents of the proton. 
The excellent agreement of the measurements with theory is rewarding, although not surprising, because the production of $\pi^0$ at large transverse momentum was discovered at the CERN-ISR in 1972 and proved that the partons of deeply inelastic scattering (DIS) interacted strongly with each other, which was explained by QCD in 1978~\cite{BBKBBGCGKS}.

\begin{figure}[!h]
\begin{center}
\includegraphics[width=0.50\linewidth]{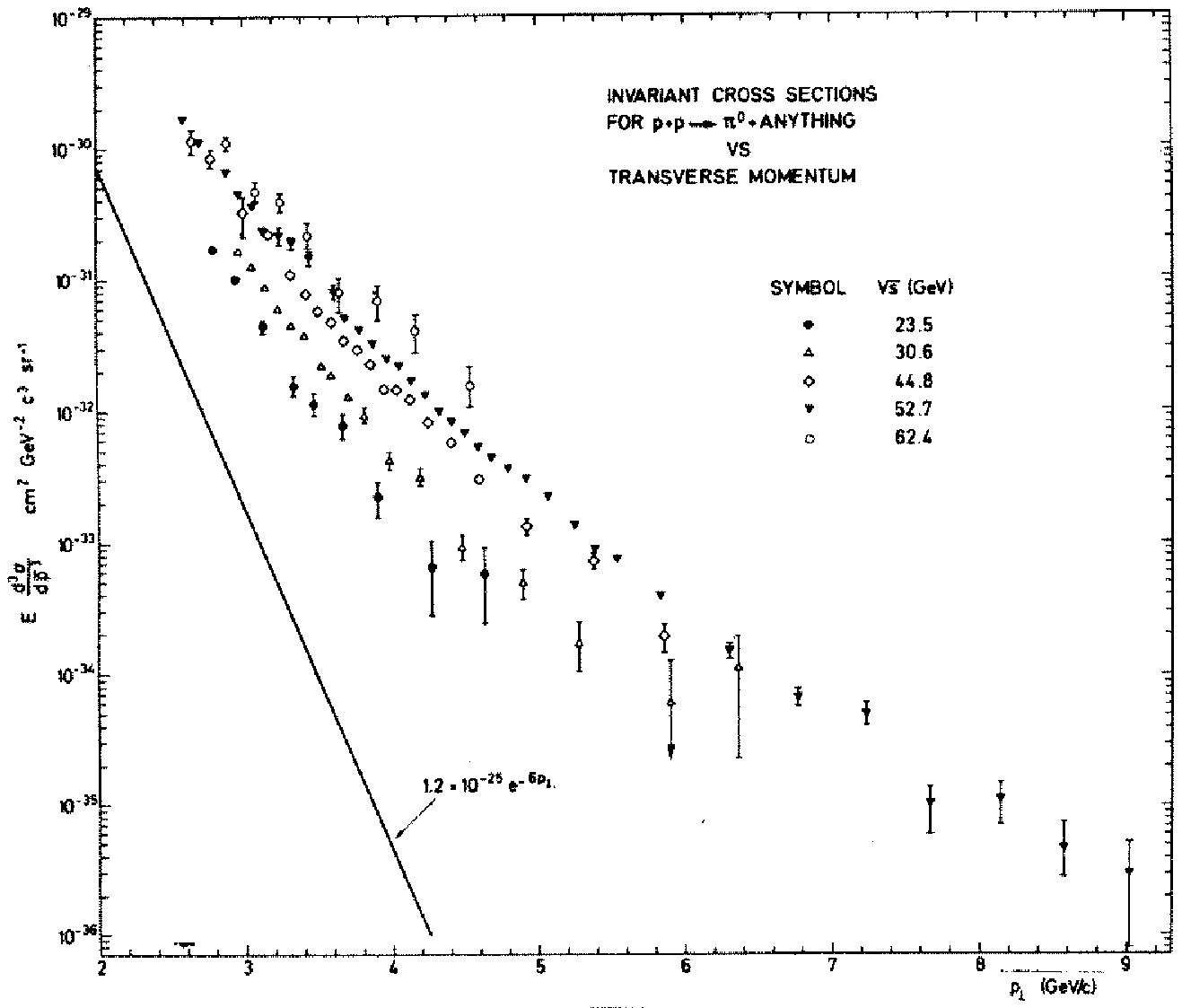}
\includegraphics[width=0.40\linewidth]{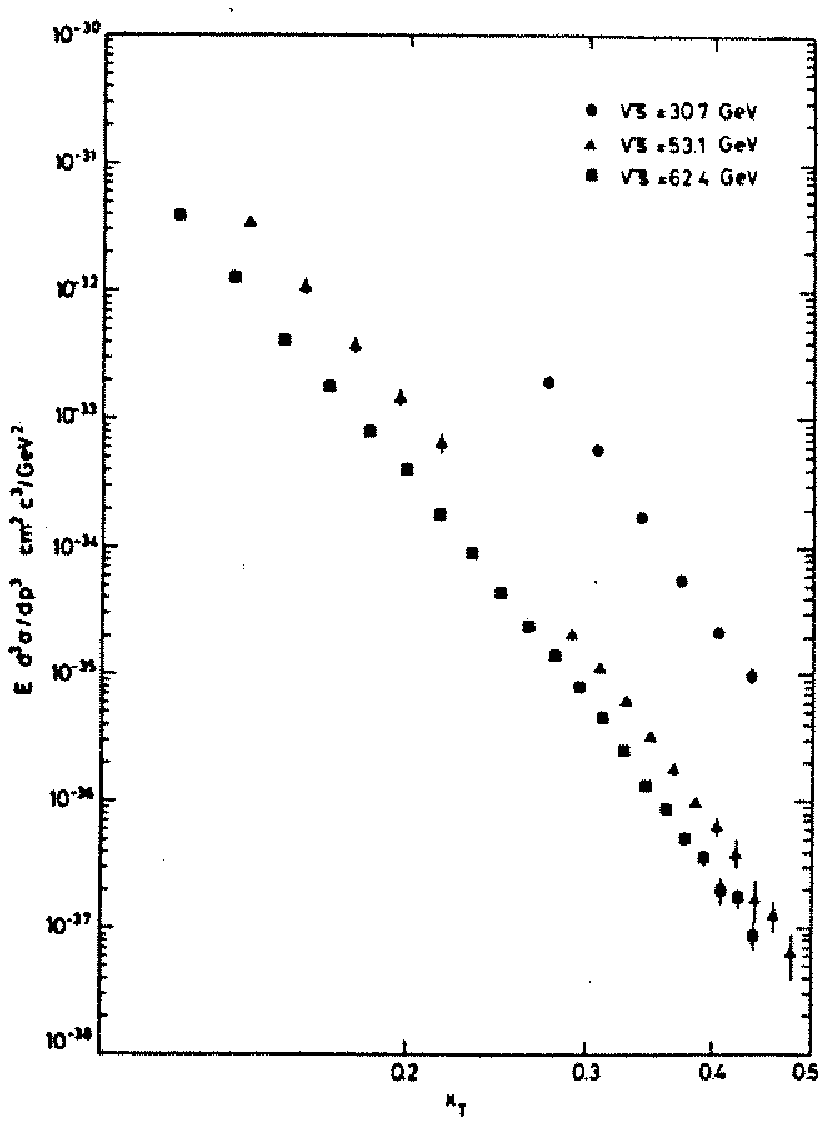}
\end{center}
\caption[]{a) (left) CCR~\cite{CCR} measurement of the invariant cross section of $\pi^0$ vs. $p_T$ at mid-rapidity in p-p collisons for 5 values of $\sqrt{s}$. b) (right) Later ISR measurement of invariant cross section of $\pi^0$ vs. $x_T=2p_T/\sqrt{s}$ at mid-rapidity in p-p collisons for 3 values of $\sqrt{s}$~\cite{CCOR}. }
\label{fig:pizeroISR}
\end{figure}

 The ISR discovery~\cite{CCR} (Fig.~\ref{fig:pizeroISR}a) showed that the $e^{-6p_T}$ dependence at low $p_T$ breaks to a power law with characteristic $\sqrt{s}$ dependence for $p_T > 2$ GeV/c, which is more evident from the log-log plot of subsequent data~\cite{CCOR} (Fig.~\ref{fig:pizeroISR}b) as a function of $x_T=2p_T/\sqrt{s}$. 
Further ISR measurements utilizing inclusive single or pairs of hadrons established that high $p_T$ particles in p-p collisions are produced from states with two roughly back-to-back jets which are  the result of scattering of constituents of the nucleons as described by Quantum Chromodynamics (QCD), which was developed during the course of those measurements~\cite{BBKBBGCGKS}. These techniques have been used extensively and further developed at RHIC since they are the only practical method to study hard-scattering and jet phenomena in Au+Au central collisions at RHIC energies. 

\begin{figure}[ht]
\begin{center}
\includegraphics[width=0.50\linewidth]{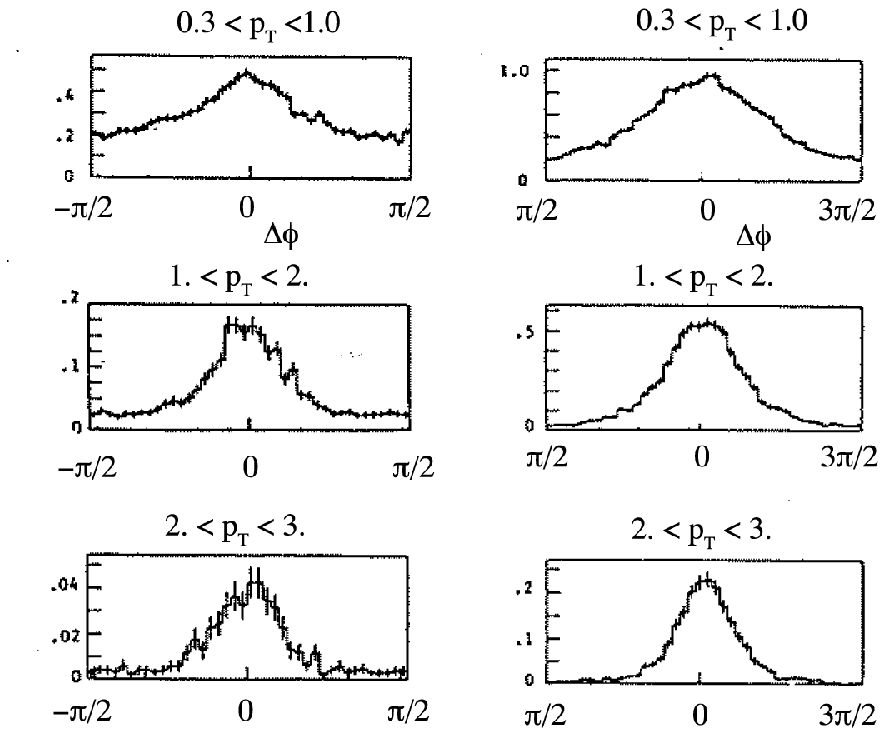} 
\hspace*{0.25in}\includegraphics[width=0.30\linewidth]{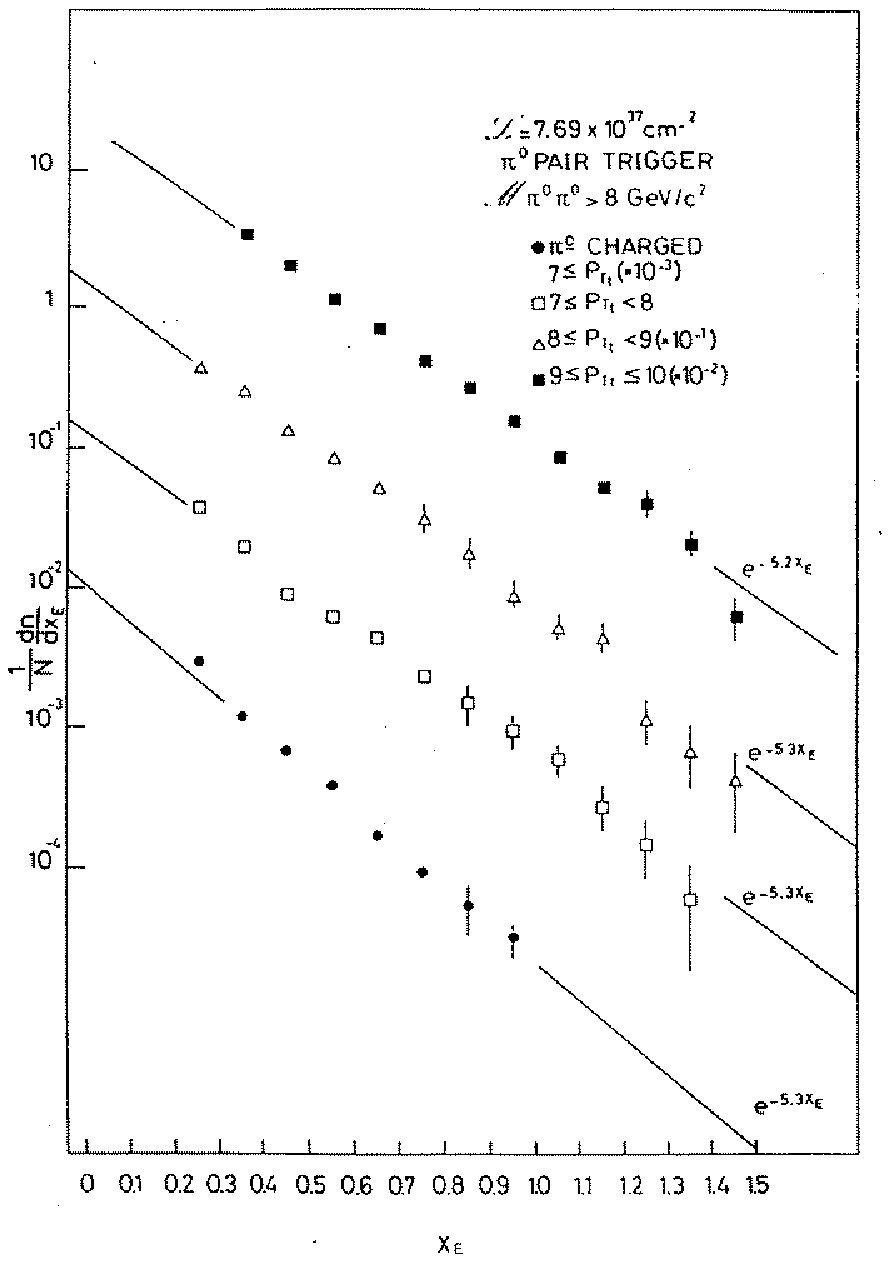}
\end{center}
\vspace*{-0.12in}
\caption[]
{a,b) Distributions of azimuthal angle ($\Delta \phi)$ of associated charged particles of transverse momentum $p_{T_a}$, with respect to a trigger $\pi^0$ with $p_{T_t}\geq 7$ GeV/c, for 3 intervals of $p_{T_{(a)}}$: a) (left-most panel) for $\Delta\phi=\pm \pi/2$ rad about the trigger particle, and b) (middle panel) for $\Delta\phi=\pm \pi/2$ about $\pi$ radians (i.e. directly opposite in azimuth) to the trigger. The trigger particle is restricted to $|\eta|<0.4$, while the associated charged particles are in the range $|\eta|\leq 0.7$. c) (right panel) $x_E$ distributions (see text) corresponding to the data of the center panel.   
\label{fig:mjt-ccorazi} }
\end{figure}

The di-jet structure of events triggered by a high $p_T$ $\pi^0$, measured via two-particle correlations, is shown in Fig~\ref{fig:mjt-ccorazi}. The peaks on both the same side (Fig.~\ref{fig:mjt-ccorazi}a) as the trigger $\pi^0$ and opposite in azimuth (Fig.~\ref{fig:mjt-ccorazi}b) are due to the correlated charged particles from jets. 
The integrated (in $\Delta\phi$) yield of the away side-particles as a function of the variable $x_E\equiv -p_{T_a} \cos(\Delta\phi)/p_{T_t}\approx z_{a}/z_{t}$, where $z_t\approx p_{T_t}/\hat{p}_{T_t}$ is the fragmentation variable of the trigger jet (with $\hat{p}_{T_t}$) and $z_a\approx p_{T_a}/\hat{p}_{T_a}$ is the fragmentation variable of the away  jet (with $\hat{p}_{T_a}$),    
was thought in the ISR era to measure the fragmentation function of the away jet (Fig.~\ref{fig:mjt-ccorazi}c) but was found at RHIC to be sensitive, instead, to the ratio of the transverse momenta of the away-jet to the trigger jet, $\hat{x}_h\equiv \hat{p}_{T_a}/\hat{p}_{T_t}$~\cite{ppg029}.

\begin{figure}[!h]
\vspace*{-1pc}
\begin{center}
\hspace*{-0.03\linewidth}\includegraphics[width=1.13\linewidth]{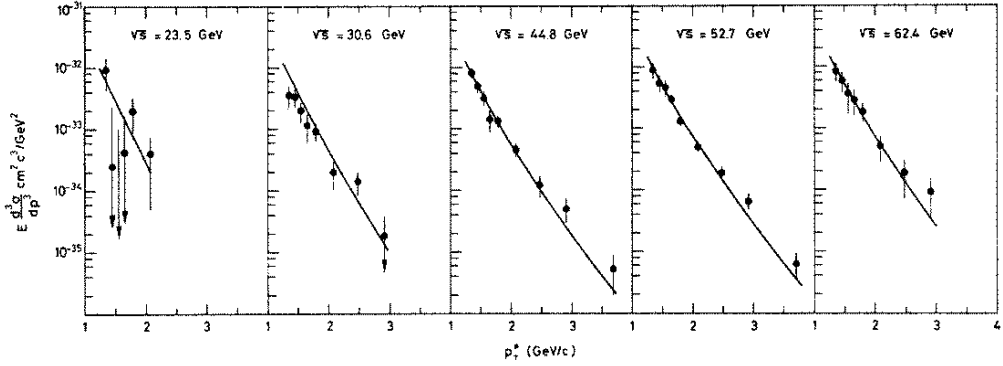}
\end{center}
\vspace*{-1.5pc}
\caption[]{Invariant cross sections at mid-rapidity at the CERN-ISR: $(e^+ + e^-)/2$ (points); $10^{-4}\times (\pi^+ +\pi^-)/2$ (lines)~\cite{CCRS}.}
\label{fig:I2}
\end{figure}

Two other ISR discoveries, direct single-$\gamma$ production and direct-single $e^{\pm}$ production, and one near miss, $J/\Psi$ production, are important components of physics at RHIC. Direct single-$\gamma$ production via the inverse QCD-compton process $g+q \rightarrow \gamma+q$ is an important probe in A+A collisions because the $\gamma$ is a direct participant in the reaction at the constituent level, emerges from the medium without interacting and can be measured precisely~\cite{egCMOR}.  Direct single-$e^{\pm}$ at a level of $e^{\pm}/\pi^{\pm}\approx 10^{\-4}$ for all values of $\sqrt{s}$ at the CERN-ISR were discovered before either the $J/\Psi$ or open-charm~\cite{CCRS} (Fig.~\ref{fig:I2}).  
After the discovery of the $J/\Psi$ in 1974, it was demonstrated that the $J/\Psi$ was not the source of the direct single-$e^{\pm}$~\cite{CCRS}  and two years later, when open charm was discovered, it was shown that the direct single-$e^{\pm}$ were due to the semi-leptonic decay of charm mesons~\cite{HLLS}. 

\section{From ISR p-p to RHIC A+A physics}

   The discovery, at RHIC, that $\pi^0$ are suppressed by roughly a factor of 5 compared to point-like scaling of hard-scattering in central Au+Au collisions is arguably {\em the}  major discovery in Relativistic Heavy Ion Physics. 
   Since hard-scattering at high $p_T >2$ GeV/c is point-like, with distance scale $1/p_T < 0.1$ fm, the cross section in p+A (A+A) collisions, compared to p-p, should be larger by the relative number of possible point-like encounters, a factor of $A$ ($A^2$) for p+A (A+A) minimum bias collisions. When the impact parameter or centrality of the collision is defined, the proportionality factor becomes $\mean{T_{AA}}$, the average overlap integral of the nuclear thickness functions. 
 
\begin{figure}[!ht]
\begin{center}
\begin{tabular}{cc}
\includegraphics[width=0.45\linewidth]{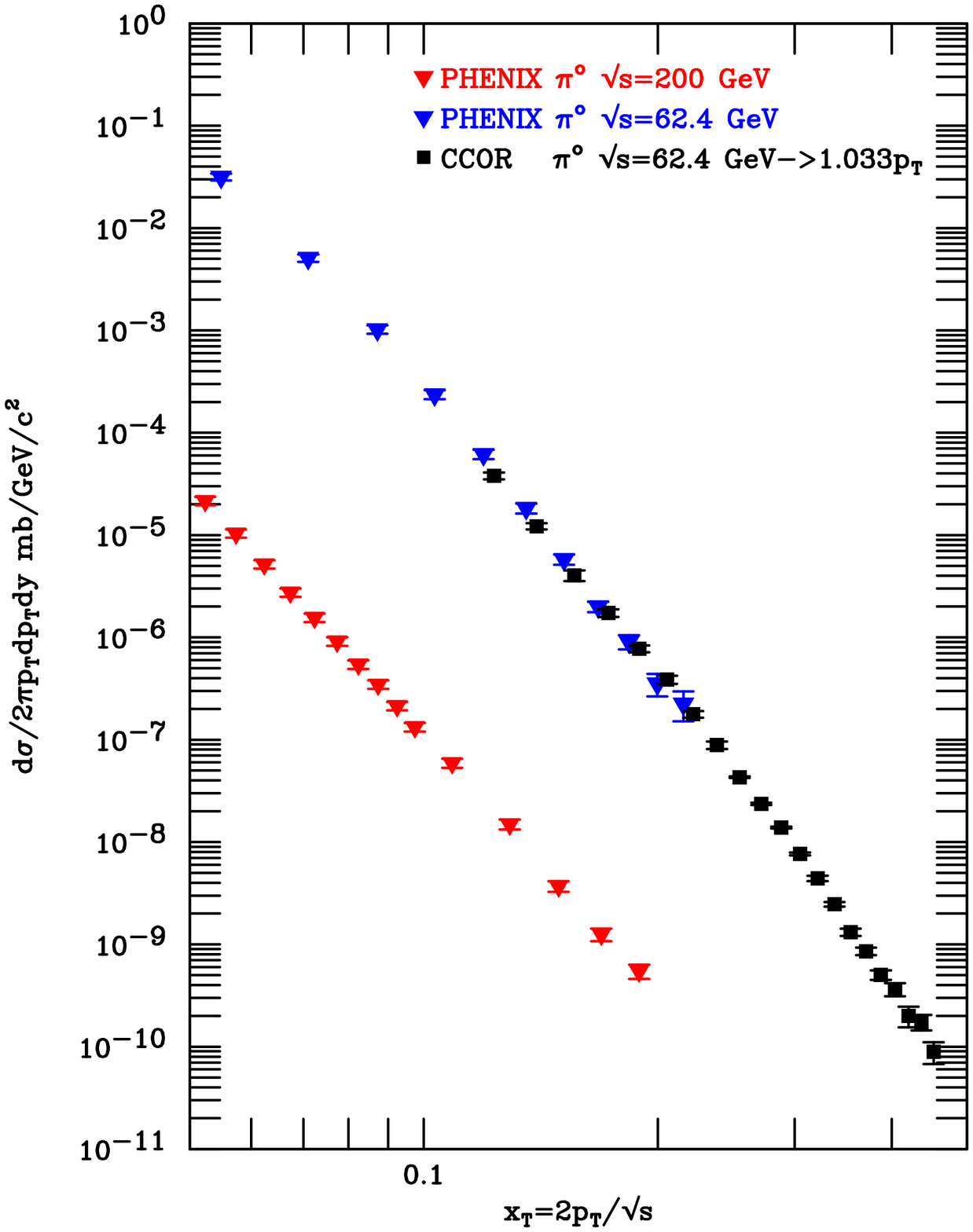} &
\hspace*{-0.04\linewidth} \includegraphics[width=0.51\linewidth]{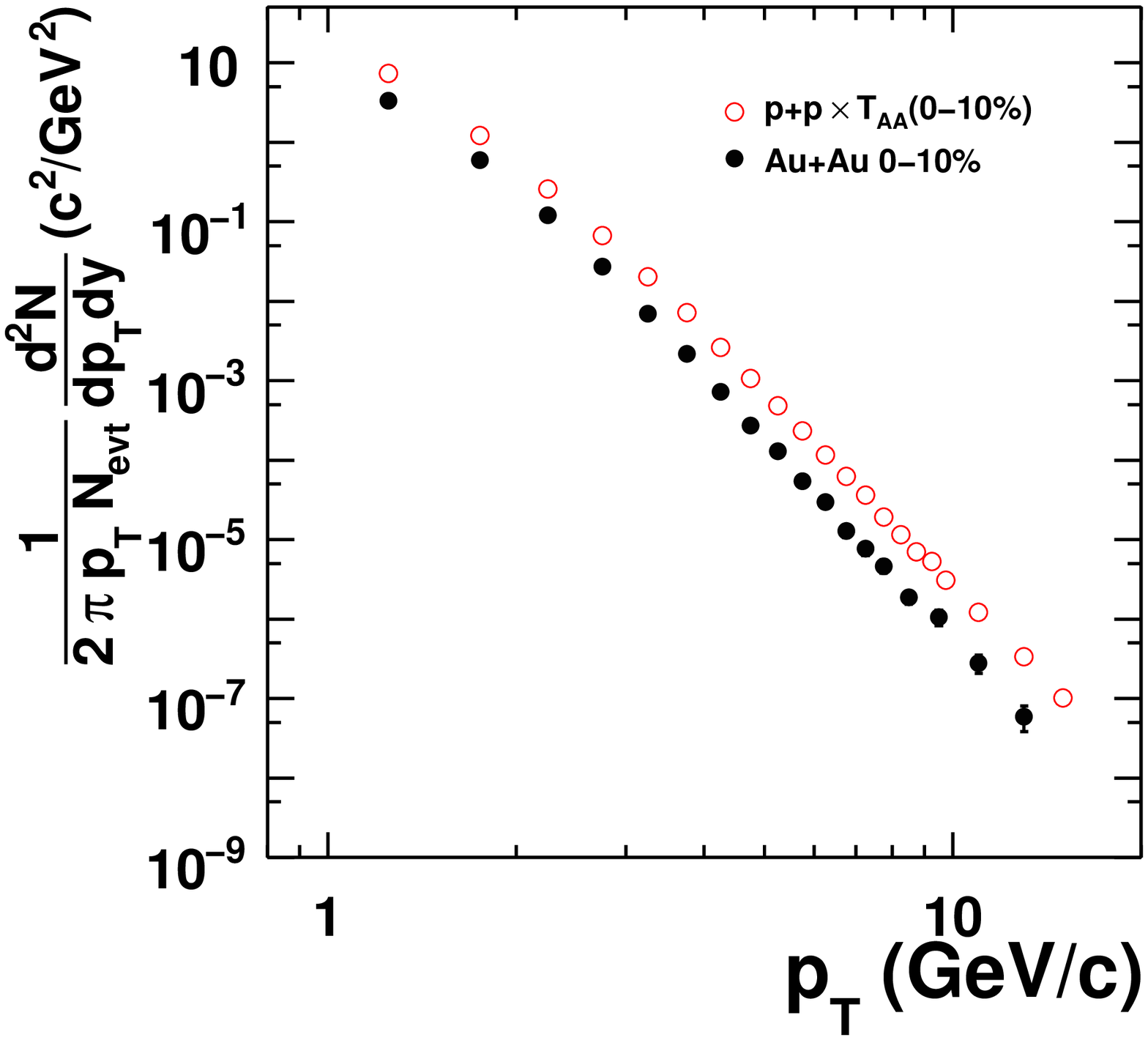}
\end{tabular}
\end{center}\vspace*{-1.5pc}
\caption[]{a) (left) $Ed^3\sigma/dp^3$ vs. $x_T$ for PHENIX mid-rapidity $\pi^0$ at $\sqrt{s}=200$ GeV in p-p collisions~\cite{PXpi0PRD} plus PHENIX and CCOR-ISR~\cite{CCOR} measurements at $\sqrt{s}=62.4$ GeV, where the absolute $p_T$ scale of the ISR measurement has been corrected upwards by 3\% to agree with the PHENIX data. b) (right) $\pi^0$ p-p data vs. $p_T$ at $\sqrt{s}=200$ GeV from (a) multiplied by $\mean{T_{AA}}$ of Au+Au central (0-10\%) collisions compared to semi-inclusive $\pi^0$ invariant yield in Au+Au central (0-10\%) collisions at $\sqrt{s_{NN}}=200$ GeV. } 
\label{fig:f1}
\end{figure}

   In Fig.~\ref{fig:f1}a), the PHENIX measurement of $E d^3 \sigma/dp^3$ for $\pi^0$ production in p-p collisions at $\sqrt{s}=62.4$ is in excellent agreement with the ISR data and the PHENIX $\pi^0$ data follow the same trend as the lower energy data, with a pure power law, $E d^3 \sigma/dp^3\propto p_T^{-8.1\pm 0.1}$ for $p_T > 3$ GeV/c at $\sqrt{s}=200$ GeV. In Fig.~\ref{fig:f1}b), the 200 GeV p-p data, multiplied by the point-like scaling factor $\mean{T_{AA}}$ for (0-10\%) central Au+Au collisions are compared to the semi-inclusive invariant $\pi^0$ yield in central (0-10\%) Au+Au collisions at $\sqrt{s_{NN}}=200$ GeV and,  amazingly, the Au+Au data follow the same power-law as the p-p data but are suppressed from the point-like scaled p-p data by a factor of $\sim 5$. The suppression is represented quantitatively by the ``nuclear modification factor'', $R_{AA}(p_T)$, the ratio of the measured semi-inclusive yield in A+A collisions to the point-like scaled p-p cross section at the same $p_T$: 
         \begin{equation}
  R_{AA}(p_T)={{d^2N^{\pi}_{AA}/dp_T dy N_{AA}}\over {\langle T_{AA}\rangle d^2\sigma^{\pi}_{pp}/dp_T dy}} \quad . 
  \label{eq:RAA}
  \end{equation}

\begin{figure}[!h]
\includegraphics[width=0.50\textwidth]{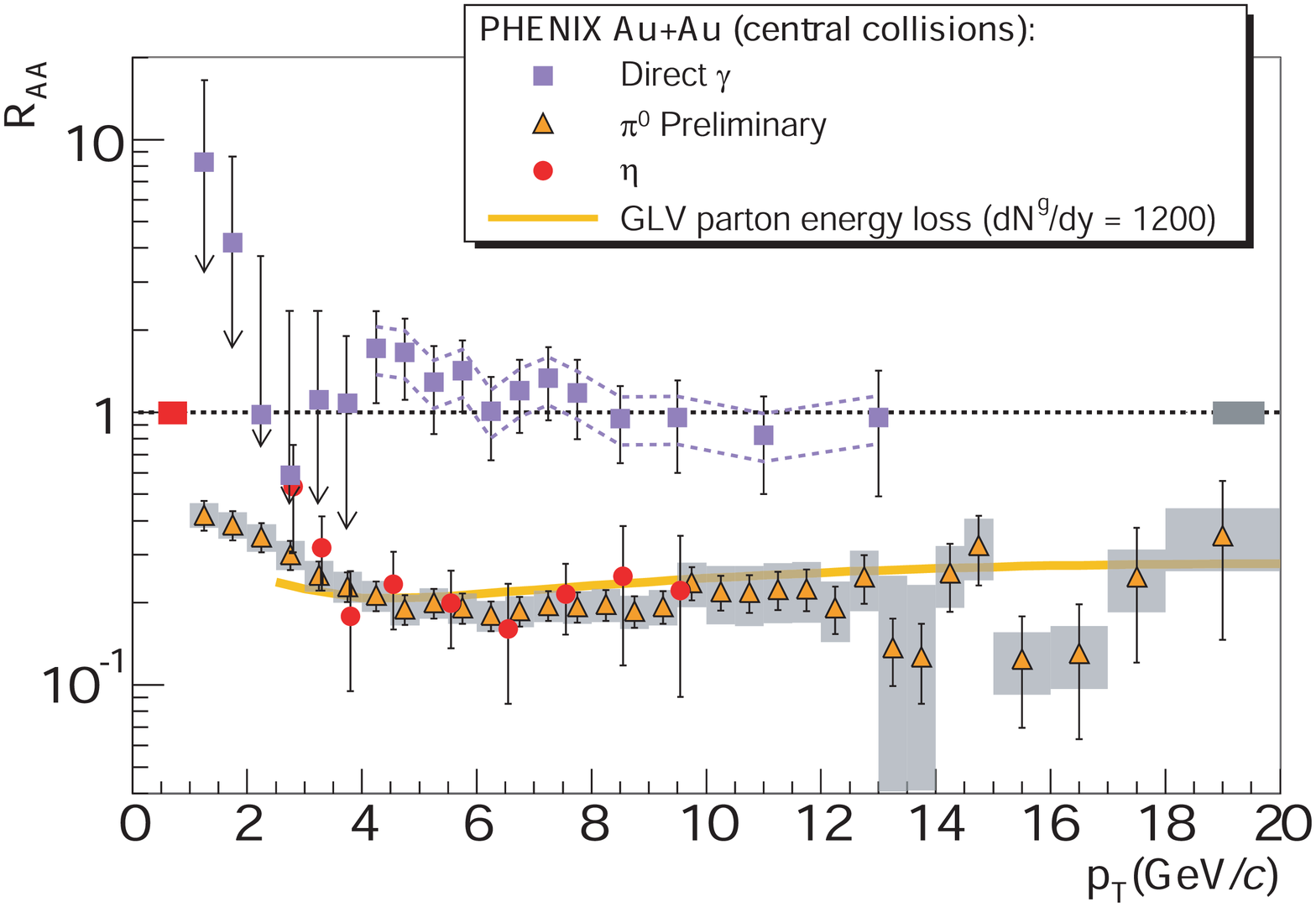} 
\includegraphics[width=0.50\textwidth]{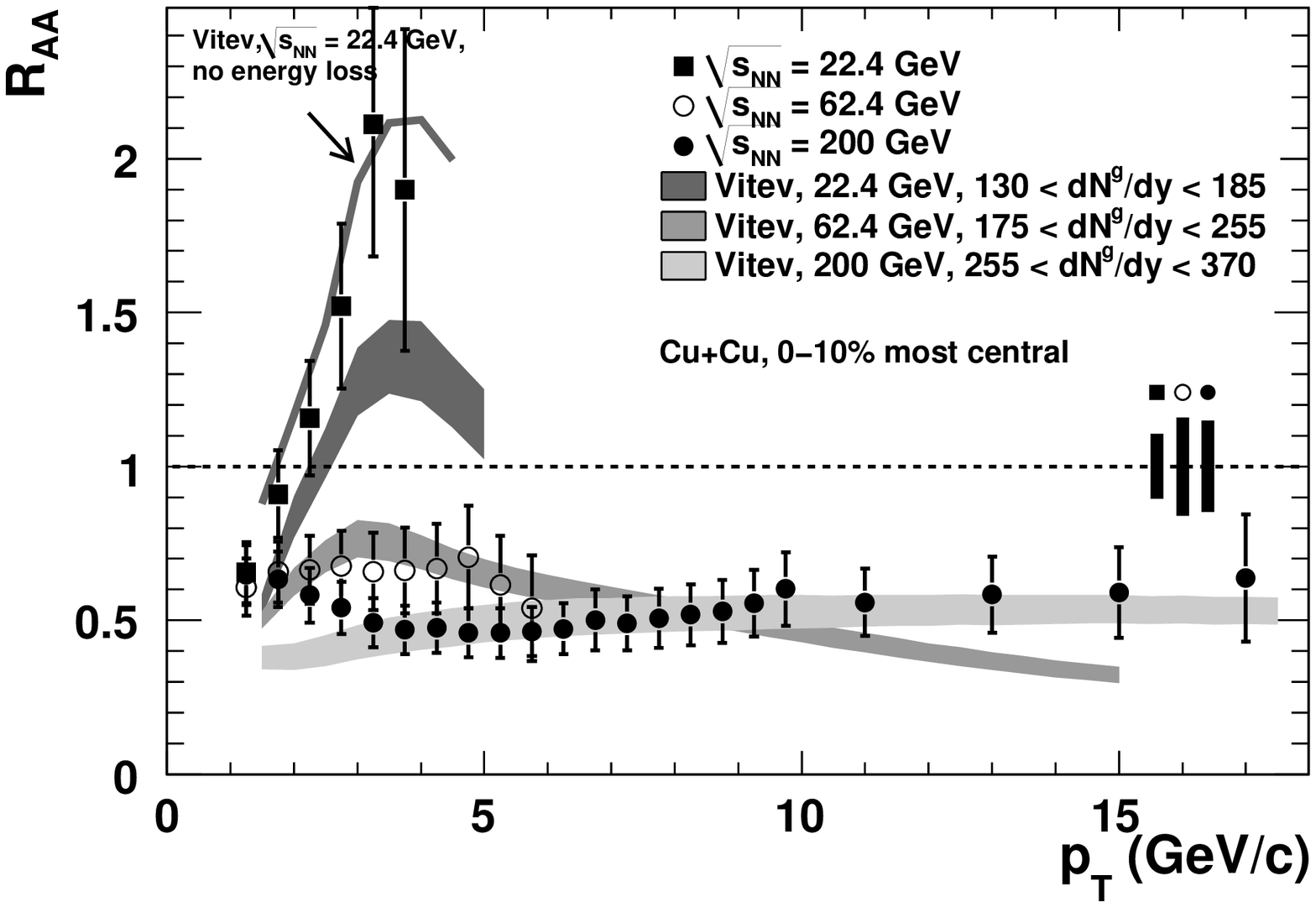} 
\caption[]{a) (left) Nuclear modification factor, $R_{AA}$ for direct-$\gamma$, $\pi^0$ and $\eta$ in Au+Au (0-10\%) central collisions at $\sqrt{s_{NN}}=200$ GeV~\cite{YAQM05}, together with GLV theory curve~\cite{GLV}. b) PHENIX $R_{AA}$ for $\pi^0$ in Cu+Cu central collisions at $\sqrt{s_{NN}}=200$, 62.4 and 22.4 GeV~\cite{ppg084}, together with Vitev theory curves~\cite{Vitev2}.}
\label{fig:QM05wowPXCu}
\end{figure}

In Fig.~\ref{fig:QM05wowPXCu}a, $R_{AA}(p_T)$ is shown for $\pi^0$, $\eta$ mesons and direct-$\gamma$ for $\sqrt{s_{NN}}=200$ GeV Au+Au central (0-10\%) collisions~\cite{YAQM05}. The $\pi^0$ and $\eta$ mesons, which are fragments of jets from outgoing partons are suppressed by the same amount while the direct-$\gamma$ which do not interact in the medium are not suppressed. This indicates a strong medium effect on outgoing partons. 

Fig.~\ref{fig:QM05wowPXCu}b shows that $R_{AA}$ for central (0-10\%) Cu+Cu collisions is comparable at $\sqrt{s_{NN}}=62.4$ and 200 GeV, but that there is no suppression, actually a Cronin enhancement, at $\sqrt{s_{NN}}=22.4$ GeV~\cite{ppg084}.  This indicates that the medium which suppresses jets is produced somewhere between $\sqrt{s_{NN}}=22.4$ GeV, the SpS Fixed Target highest c.m. energy, and 62.4 GeV. 

	Remarkably, the dramatic difference in $\pi^0$ suppression from SpS to RHIC is not reflected in $J/\Psi$ suppression, which is nearly identical at mid-rapidity at RHIC and the SpS, casting some doubt on the value of $J/\Psi$ suppression as a probe of deconfinement~\cite{MJTROP}. One possible explanation is that $c$ and $\bar{c}$ quarks in the QGP recombine to regenerate $J/\Psi$,  miraculously making the observed $R_{AA}$ equal at SpS and RHIC c.m. energies. The good news is that such models predict $J/\Psi$ enhancement ($R_{AA}> 1$) at LHC energies, which would be spectacular, if observed.       

	Returning to the jet suppression shown in Fig.~\ref{fig:QM05wowPXCu}, the measurements seem to be in excellent agreement with the theoretical curves~\cite{GLV,Vitev2}. The suppression can be explained by the energy loss of the outgoing partons in the dense color-charged medium due to coherent Landau-Pomeranchuk-Migdal radiation of gluons, predicted in QCD~\cite{BDMPS}, which is sensitive properties of the medium. 
	  \begin{figure}[!ht]
\begin{center}
\begin{tabular}{cc}
\begin{tabular}[b]{c}
\hspace*{-0.03\linewidth}\includegraphics[width=0.509\linewidth]{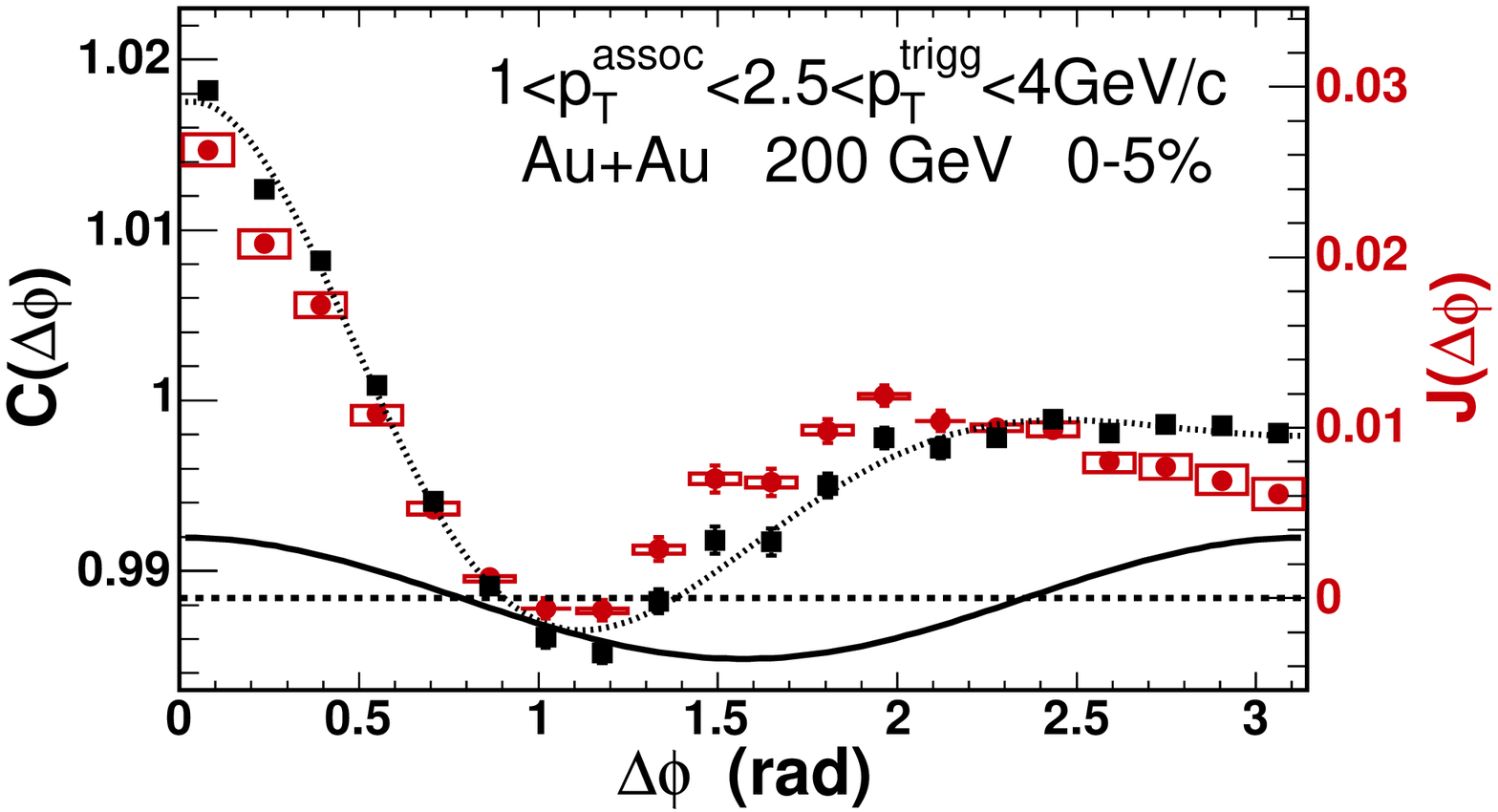}\cr
\hspace*{-0.03\linewidth}\includegraphics[width=0.435\linewidth]{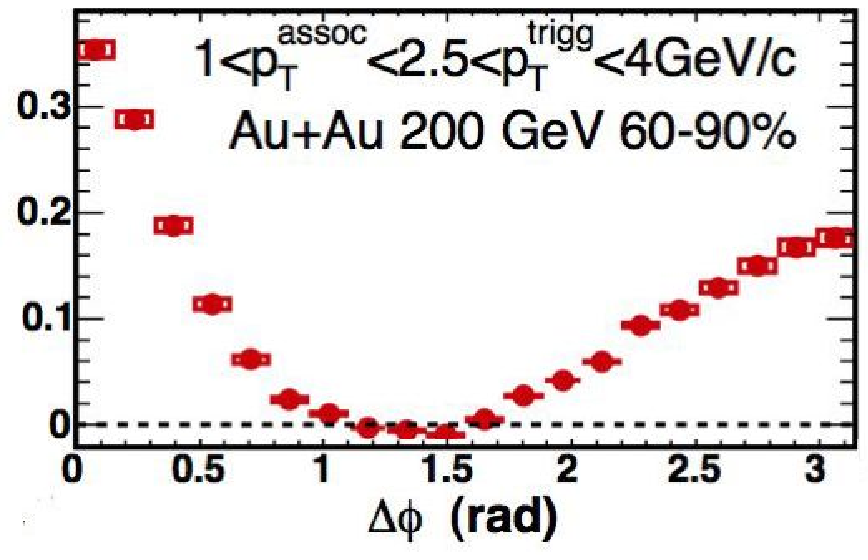}\hspace*{0.03\linewidth}
\end{tabular} 
\hspace*{-0.01\linewidth}\includegraphics[width=0.471\linewidth]{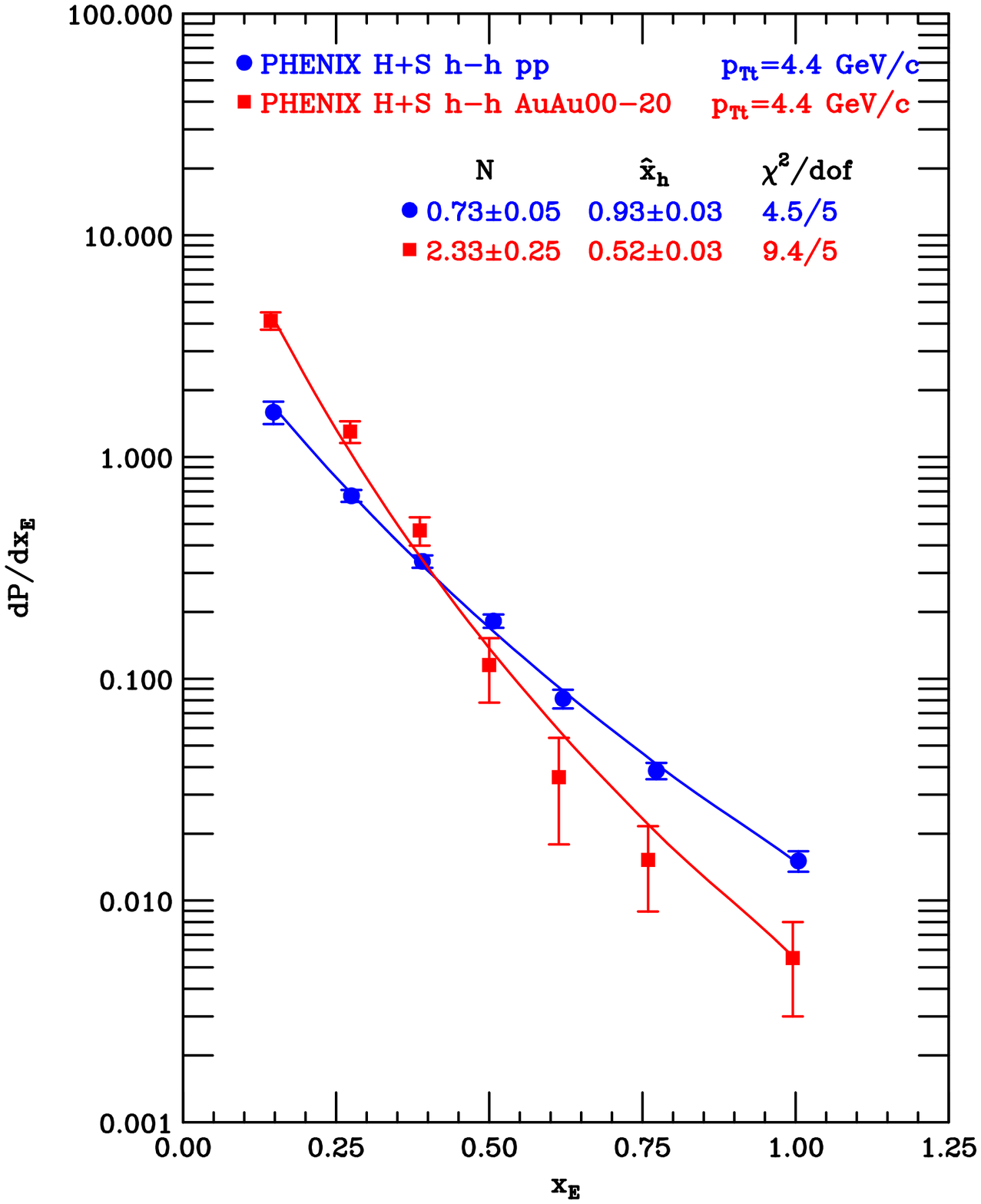}
\end{tabular}
\end{center}\vspace*{-1.7pc}
\caption[]{a) (left) Conditional yield azimuthal correlation function, $C(\Delta\phi)$ (black squares), flow background  (solid line) and Jet function $J(\Delta\phi)$ (red dots) after flow subtraction, per trigger $h^{\pm}$ with $2.5< p_{T_t}<4$ GeV/c for associated $h^{\pm}$ of $1.0 < p_{T_a}<2.5$ GeV/c from PHENIX~\cite{ppg067}. b) (right) $x_E\approx p_{T_a}/p_{T_t}$ distribution for the Au+Au-central data compared to p-p. }
\label{fig:f8}
\end{figure}
	Measurements of two-particle correlations confirm the loss of energy of the away-jet relative to the trigger jet in Au+Au central collisions compared to p-p collisions. In analogy to Fig.~\ref{fig:mjt-ccorazi} (above), the two-particle correlations in Au+Au collisions (Fig.~\ref{fig:f8}a) show clear di-jet structure in both peripheral and central collisions. The away-side correlation in central Au+Au collisions is much wider than in peripheral Au+Au and p-p collisions and is further complicated by the large multiparticle background which is a modulated in azimuth by the $v_2$ collective flow of a comparable width to the jet correlation. After the $v_2$ correction, a double peak structure $\sim \pm 1$ radian from $\pi$ is evident, with a dip at $\pi$ radians. This may indicate a reaction of the medium to a passing parton in analogy to a ``sonic-boom'' and is under active study both theoretically and experimentally. The energy loss of the away-parton is indicated by the fact that the $x_E$ distribution in Au+Au central collisions (Fig.~\ref{fig:f8}b) is steeper than that from p-p collisions. As noted above,   
we found in PHENIX that the $x_E$ distribution did not measure the fragmentation function of the away-jet but is sensitive instead to $\hat{x}_h$, the ratio of the transverse momentum of the away-parton to that of the trigger parton, specifically~\cite{ppg029}:
\begin{equation}
\left . {dP\over dx_E}\right |_{p_{T_t}}=N (n-1) {1\over \hat{x}_h} {1 \over{(1+x_E/\hat{x}_h})^n}
\label{eq:xEdist} 
 \end{equation}
 where $N$ is a normalization factor, and $n$ (=8.1 at 200 GeV) is the power of the inclusive invariant $p_{T_t}$ distribution. The value of $\hat{x}_h=0.52$ in Au+Au collisions compared to 0.93 in p-p collisions shows that the away-parton in Au+Au has lost roughly 1/2 its energy before fragmenting.

\subsection{A theoretical crisis-heavy quark suppression}
\begin{figure}[!ht]
\begin{center} 
\begin{tabular}{cc}
\hspace*{-0.04\linewidth}\includegraphics*[width=0.53\linewidth]{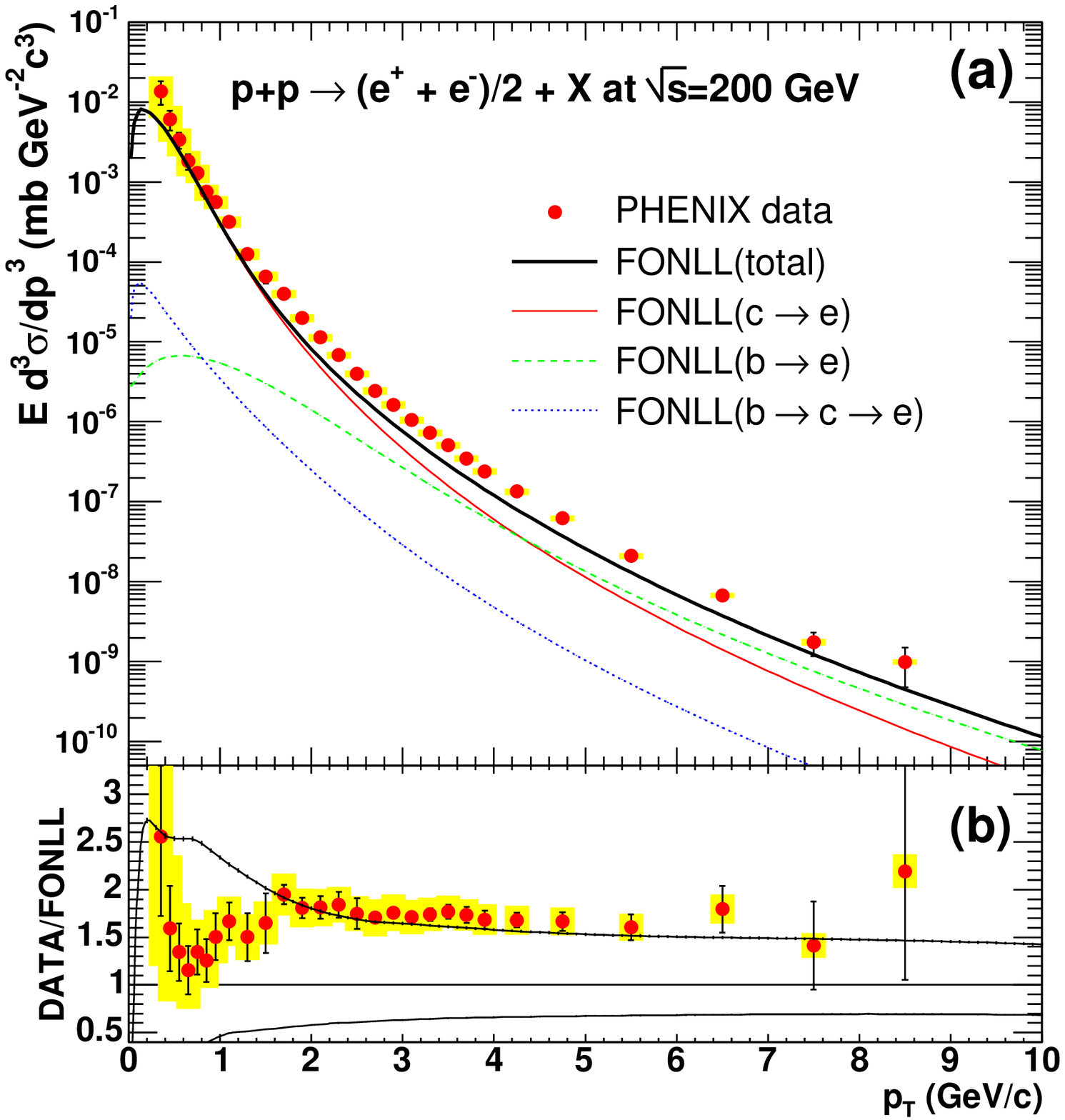} & 
\hspace*{-0.02\linewidth}\includegraphics*[width=0.51\linewidth]{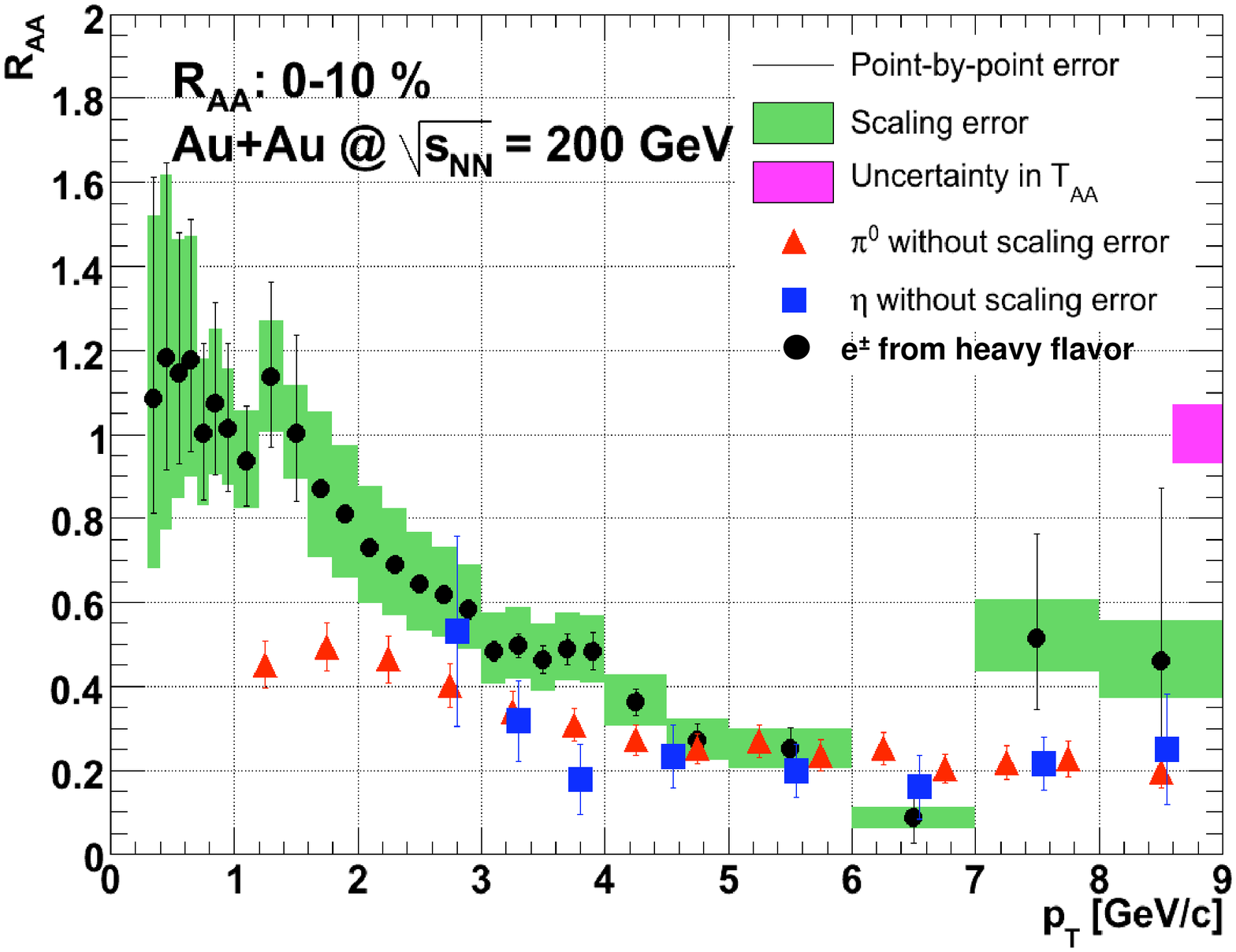} 
\end{tabular}
\end{center}\vspace*{-1.5pc}
\caption[]{a) (left) Invariant cross section of direct single-$e^{\pm}$ in p-p collisions ~\cite{PXcharmpp06} compared to theoretical predictions from $c$ and $b$ quark semileptonic decay. b) (right) $R_{AA}$ as a function of $p_T$ for direct single-$e^{\pm}$~\cite{PXPRL97e}, $\pi^0$ and $\eta$ in Au+Au central (0-10\%) collisions at $\sqrt{s_{NN}}=200$ GeV.}
\label{fig:f7}
\end{figure}

   PHENIX was specifically designed to be able to detect charm particles via direct single-$e^{\pm}$. Fig.~\ref{fig:f7}a shows our direct single-$e^{\pm}$ measurement in p-p collisions at $\sqrt{s}=200$ GeV~\cite{PXcharmpp06} in agreement with a QCD calculation of $c$ and $b$ quarks as the source of the direct single-$e^{\pm}$ (also called non-photonic $e^{\pm}$ at RHIC).   

In Au+Au collisions, a totally unexpected result was observed. The direct single-$e^{\pm}$ from heavy quarks are suppressed the same as the $\pi^0$ and $\eta$ from light quarks (and gluons) in the range $4\leq p_T\leq 9$ GeV/c where $b$ and $c$ contributions are roughly equal (Fig.~\ref{fig:f7}b)~\cite{PXPRL97e}. This strongly disfavors the QCD energy-loss explanation of jet-quenching because, naively, heavy quarks should radiate much less than light quarks and gluons in the medium; but opens up a whole range of new possibilities including string theory~\cite{egsee066}. 
\section{Zichichi to the rescue?}
  In September 2007, I read an article by Nino, ``Yukawa's gold mine'', in the CERN Courier taken from his talk at the 2007 International Nuclear Physics meeting in Tokyo, Japan, in which he proposed: ``the reason why the top quark appears to be so heavy (around 200 GeV) could be the result of some, so far unknown, condition related to the fact that the final
state must be QCD-colourless. We know that confinement produces masses of the order of a giga-electron-volt. Therefore, according to our present understanding, the QCD colourless condition cannot explain the heavy quark mass. However, since the origin of the quark masses is still not known, it cannot be excluded that in a QCD coloured world, the six quarks are all nearly massless and that the colourless condition is `flavour' dependent.'' 
  
  Nino's idea really excited me even though, or perhaps because, it appeared to overturn two of the major tenets of the Standard Model since it seemed to imply that: QCD isn't flavor blind;  the masses of quarks aren't given by the Higgs mechanism.  Massless $b$ and $c$ quarks in a color-charged medium would be the simplest way to explain the apparent equality of gluon, light quark and heavy quark suppression indicated by the equality of $R_{AA}$ for $\pi^0$ and $R_{AA}$ of direct single-$e^{\pm}$ in regions where both $c$ and $b$ quarks dominate. Furthermore RHIC and LHC-Ions are the only place in the Universe to test this idea. 
  
   It may seem surprising that I would be so quick to take Nino's idea so seriously. This confidence dates from my graduate student days when I checked the proceedings of the 12th ICHEP in Dubna, Russia in 1964 to see how my thesis results were reported and I found several interesting questions and comments by an ``A. Zichichi'' printed in the proceedings. One comment about how to find the $W$ boson in p+p collisions deserves a verbatim quote because it was exactly how the $W$ was discovered at CERN 19 years later: ``We would observe the $\mu$'s from W-decays. By measuring the angular and momentum distribution at large angles of K and $\pi$'s, we can predict the corresponding $\mu$-spectrum. We then see if the $\mu$'s found at large angles agree with or exceed the expected numbers.''
   
  Nino's idea seems much more reasonable to me than the string theory explanations of heavy-quark suppression (especially since they can't explain light-quark suppression). Nevertheless, just to be safe, I asked some distinguished theorists what they thought, with these results:
  ``Oh, you mean the Higgs field can't penetrate the QGP'' (Stan Brodsky);
``You mean that the propagation of heavy and light quarks through the medium is the same'' (Rob Pisarski); 
``The Higgs coupling to vector bosons $\gamma$, $W$, $Z$ is specified in the standard model and is a fundamental issue. One big question to be answered by the LHC is whether the Higgs gives mass to fermions or only to gauge bosons. The Yukawa couplings to fermions are put in by hand and are not required'' ``What sets fermion masses, mixings?" (Chris Quigg-Moriond2008); ``No change in the $t$-quark, $W$, Higgs mass relationship   if there is no Yukawa coupling: but there could be other changes'' (Bill Marciano).

	 Nino proposed to test his idea by shooting a proton beam through a QGP formed in a Pb+Pb collision at the LHC and seeing the proton `dissolved' by the QGP. My idea is to use the new PHENIX vertex detector, to be installed in 2010, to  map out, on an event-by-event basis, the di-hadron correlations from identified $b,\overline{b}$ di-jets, identified $c,\bar{c}$ di-jets, which do not originate from the vertex, and light quark and gluon di-jets, which originate from the vertex and can be measured with $\pi^0$-hadron correlations. A steepening of the slope of the $x_E$ distribution of heavy-quark correlations as in Fig.~\ref{fig:f8}b will confirm in detail (or falsify) whether the different flavors of quarks behave as if they have the same mass in a color-charged medium. If Nino's proposed effect is true, that the masses of fermions are not given by the Higgs, and we can confirm the effect at RHIC or LHC-Ions, this would be a case where Relativistic Heavy Ion Physics may have something unique to contribute at the most fundamental level to the Standard Model, which would constitute a ``transformational discovery.'' Of course the LHC could falsify this idea by finding the Higgs decay to $b,\overline{b}$ at the expected rate in p-p collisions. Clearly, there are exciting years ahead of us!

\end{document}